\documentstyle[aps,12pt]{revtex}

\begin{document}
\title{Role of f electrons in rare-earth and uranium intermetallics\\
-An alternative look at heavy-fermion phenomena.}
\author{R.J. Radwanski}
\address{Centre for Solid State Physics, Filipa 5, 31-150 Krak\'{o}w, POLAND}
\maketitle

The origin of the large specific heat and the non-magnetic state observed at
low temperatures in some f intermetallic compounds with uranium called
heavy-fermion (h-f) compounds is discussed. Different existing theoretical
models are briefly overviewed but it will be proposed to discuss the h-f
compounds in terms of physical concepts worked out for rare-earth
intermetallics [1]. To remind, the magnetic and electronic properties of
rare-earth intermetallics are understood by considering a few, two in the
simplest but quite adequate approach, electronic subsystems i.e. the f
electronic subsystem and conduction-electron subsystem (the
individualized-electron model). These two subsystems are described by
essentially different theoretical approaches referring to the localized and
band magnetism. 

In the discussion if the non-magnetic state observed in h-f compounds do
refer to the local scale (single-ion) or to a collective many-body state
some arguments will be given for the on-site effect. Namely, it can be
rigorously proven that charge interactions via the Stark effect can produce
the non-magnetic state of the localized $f^{\text{ }n}$ electronic subsystem
also in case of the Kramers system (n is an odd number) [2]. The full
suppression of the local moment is attained by highly anisotropic charge
distribution at the vicinity of the f-shell electrons. This highly
anisotropic charge distribution is visualized by CEF parameters with
significant values for higher-order terms. The charge mechanism for the
formation of the non-magnetic state of the f magnetic ion is compared with
other known mechanism like the Kondo-compensation mechanism (the spin type)
and the hybridization, of f and conduction electrons, mechanism. 

In view of the individualized-electron model the large specific heat
originates from low-energy excitations between doublet levels of the Kramers
state of the fn electronic subsystems that are slightly split due to
exchange interactions. These excitations are many-electron excitations in
contrary to single-electron excitations in the conduction-electron
subsystem. It will be shown that magnetic and electronic properties of
intermetallic systems with the f-electronic subsystem in a quasi-nonmagnetic
Kramers state exhibit properties observed in h-f compounds. One can say that
the h-f compounds are compounds with Kramers f ions that have difficulties,
due to exotic ground state and weakness of exchange interactions, to form
the well-established magnetic order. However, the system has to release the
Kramers entropy before reaching zero temperature as is experimentally
observed by the entropy of R ln2. 

These phenomena will be discussed for some uranium h-f compounds with the
hexagonal symmetry. For instance, the temperature dependence of the specific
heat of UPd$_2$Al$_3$ with a $\lambda $-type of peak at T$_N$ of 14 K and a
Schottky-type of peak above TN has been very well reproduced by the U$^{3+}$
(5$f$ $^3$) configuration [3].

Prezented at European Conf. Physics of Magnetism 93, 21-24 June 1993,
Poznan, Poland (Committee: J.Morkowski, R.Micnas, S.Krompiewski)

\end{document}